\documentclass[10pt]{revtex4}
\usepackage{amssymb}
\usepackage{latexsym}
\usepackage{epsfig}
\begin{document}

\title{Analytical study of holographic superconductor in Born-Infeld electrodynamics with backreaction}
\author{A. Sheykhi$^{1,2}$\footnote{asheykhi@shirazu.ac.ir} and F. Shaker$^{1}$}
\address{$^1$ Physics Department and Biruni Observatory, College of
Sciences, Shiraz University, Shiraz 71454, Iran\\
$^2$ Research Institute for Astronomy and Astrophysics of Maragha
(RIAAM), P.O. Box 55134-441, Maragha, Iran}

\begin{abstract}

We generalize the analytical investigation on the properties of
$s$-wave holographic superconductors in the presence of
Born-Infeld nonlinear electrodynamics by taking the backreaction
into account. We find that in the presence of nonlinear gauge
field, one can still employ the analytical method when the
backreaction is turned on. Our calculation is based on the
Sturm-Liouville eigenvalue problem, which is a variational method.
For the system under consideration, we obtain the relation between
the critical temperature and the charge density. We find that both
backreaction and Born-Infeld parameters decrease the critical
temperature of the superconductor and make the condensation
harder. At the ned of paper, we calculate the critical exponent
associated with the condensation near the critical temperature and
find that  it has the universal value $1/2$ of the mean field
theory.

\end{abstract}

 \maketitle

\section{Introduction}
The idea of gauge/gravity duality \cite{LB} for analyzing the
real-world has been one of the remarkable achievements which
emerges from development in string theory. It has been
well-established that such a duality can provides a novel method
for calculating the properties of the superconductors by using a
dual classical gravity description. In particular, it can shed
some light on the unsolved mysteries in modern condensed matter
physics, namely, the mechanism of the high temperature
superconductors. The investigation on this subject has got a lot
of enthusiasm in the past decades \cite{Mal,Gub,Wit}.

A great step in this direction was put forwarded by Hartnoll, et.
al., \cite{Har,Har2} who disclosed that some properties of
strongly coupled superconductors can be potentially described by
classical general relativity living in one higher dimension. This
novel idea is usually called \textit{holographic superconductors}.
The holographic $s$-wave superconductor model known as
Abelian-Higgs model was first realized in \cite{Har,Har2}.
According to the anti-de Sitter/conformal field theories (AdS/CFT)
correspondence, in the gravity side, a Maxwell field and a charged
scalar field are introduced to describe the $U(1)$ symmetry and
the scalar operator in the dual field theory, respectively. This
holographic model undergoes a phase transition from black hole
with no hair (normal phase/conductor phase) to the case with
scalar hair at low temperatures (superconducting phase). Following
\cite{Har,Har2}, an overwhelming fluids of papers have been
appeared which try to apply this novel idea for understanding the
strongly coupled holographic superconductors in different setups
\cite{P.GWWY, P.H, P.G, P.BGRL, P.MRM, P.CW, P.ZGJZ,RGC2,Roy,Sub}.
For a review on the holographic superconductors see
\cite{Hor,Mus,RGC}.

On the other hand, there have been a lot of interest in studying
the high order correction terms related to the gauge field, in the
holographic superconductors. The motivation is to investigate the
effects of the nonlinear electrodynamics on the scalar
condensation. The effects of Born-Infeld nonlinear electrodynamics
on the holographic superconductors has been studied numerically in
\cite{P.JC}. Based on the Sturm-Liouville eigenvalue problem,
several properties of holographic $s$-wave superconductors in the
background of a Schwarzschild-AdS spacetime and in the presence of
Born-Infeld nonlinear electrodynamic were analytically studied in
\cite{Gan}. Similar studies were also done when the nonlinear
electrodynamics is in the form of power-Maxwell field
\cite{P.JPC}. It was shown that the larger power parameter $q$ for
the power-Maxwell field makes it harder for the scalar hair to be
condensated \cite{P.JPC}. Other studies on the holographic
superconductors in the presence of nonlinear electrodynamics were
carried out in \cite{Jing,LPJW2015}.

It is worth noting that most investigations on the holographic
superconductors focus on the probe limit in which the scalar and
gauge field do not back react on the metric background. Another
interesting behavior is when the backreaction of the matter fields
on the background geometry is taken into account. This might bring
rich physics in the holographic model away from the probe limit.
It was shown that when the backreaction is taken into account,
even the uncharged scalar field can form a condensate in the
$(2+1)$-dimensional holographic superconductor model \cite{Har2}.
Analytical and numerical investigations, based on the both
matching and Sturm-Liouville method, have been carried out for
calculating the critical temperature of the holographic
superconductor and other physical quantities when the
backreactions is turn on \cite{b.PW, b.GL,
G12013,b.KNQ,b.BH,b.PKLW,b.YJ}. Employing the variational method
for the Sturm-Liouville eigenvalue problem, the properties of the
holographic superconductors with backreaction and with linear
Maxwell field were investigated analytically in \cite{PJWC2012}.

In this paper, we would like to extend the investigations on the
holographic superconductors with backreaction by replacing the
linear Maxwell field with the nonlinear Born-Infeld
electrodynamics. We shall employ the Sturm-Liouville eigenvalue
problem to analytically investigate the properties of these
holographic superconductors. We find the critical temperature and
critical exponent of the holographic superconductor in the
presence of Born-Infeld electrodynamics with backreaction. This
eventually helps us to consider the strength of both Born-Infeld
and backreaction parameters on the condensation of the holographic
supercondoctor.

This paper is structured as follows. In section \ref{Basic}, we
present action and basic field equations of the holographic
superconductors when the Born-Infeld field and scalar field
backreact on the metric. In section \ref{Field}, we analytically
investigate the properties of these holographic superconductors by
using the Sturm-Liouville method and obtain the critical
temperature and charge density. In section \ref{Cri}, we calculate
the critical exponent and the condensation values of the
holographic superconductor. We summarize our results in section
\ref{Con}.
\section{Basic Equations of Holographic Superconductors with Backreactions \label{Basic}}
Our starting point is the following action in which gravity is
coupled to a charged, complex scalar field and Born-Infeld
nonlinear electrodynamics,
\begin{eqnarray}\label{action}
S=\int d^{4}x \sqrt{-g}\left[ \frac{1}{2 \kappa^{2}}( R-2 \Lambda)
+L_{BI}- | \nabla\psi - i q A \psi|^{2} - m^{2} |\psi|^{2}\right],
\end{eqnarray}
where $\kappa^2=8\pi G_{4}$ is the $4$-dimensional gravitational
constant, $\Lambda=-{3}/{L^2}$ is the cosmological constant and
$R$ is the Ricci scalar. Here $A$ and $\psi$ are, respectively,
the gauge and scalar field with charge $q$. The Lagrangian density
of the Born-Infeld electrodynamics is defined as
\begin{eqnarray}
L_{\rm{BI}}&=&\frac{1}{b}\Bigg(1-\sqrt{1+\frac{b F}{2}}\,\Bigg),
\end{eqnarray}
where $F=F_{\mu\nu}F^{\mu\nu}$ and $F^{\mu\nu}$ is the
electromagnetic field tensor. The constant $b$ is the Born-Infeld
coupling parameter which indicates the strength of the
nonlinearity. When $b\rightarrow0$ the Lagrangian of the
Born-Infeld reduces to the standard Maxwell Lagrangian,
$L_M=-\frac{1}{4}F_{\mu\nu}F^{\mu\nu}$. Varying the action
(\ref{action}) with respect to the metric, scalar and
electrodynamic fields leads to the following field equations
\begin{eqnarray}\label{Eq1}
&&R^{\mu\nu}-\frac{g^{\mu\nu}}{2} R -\frac{3 g^{\mu\nu}}{L^2}=
\frac{\kappa^2}{b} g^{\mu\nu} \Bigg(1-\sqrt{1+\frac{b
F}{2}}\,\Bigg)+ \frac{\kappa^2}{\sqrt{1+\frac{b F}{2}}} F_{\sigma}
^{\mu} F^{\sigma\nu}\nonumber\\
&& - \kappa^2 g^{\mu\nu} m^2 \psi^2- \kappa^2 g^{\mu\nu} |
\nabla\psi - i q A \psi|^{2}+\kappa^2 \left[(\nabla^{\nu}+ i q
A^{\nu}) \psi^{*}  (\nabla^{\mu}- i q A^{\mu})\psi+\mu
\leftrightarrow\nu\right],
\end{eqnarray}
\begin{eqnarray}\label{Eq2}
(\nabla_{\mu} -i q A_{\mu}) (\nabla^{\mu}- i q A^{\mu}) \psi -m^2
\psi =0,
\end{eqnarray}
\begin{eqnarray}\label{Eq3}
 \nabla_{\nu} \Bigg (\frac{F^{\nu\mu}}{\sqrt{1+\frac{b F}{2}}}\Bigg)= i q \Bigg [ \psi^{*} (\nabla^{\mu}- i q A^{\mu})
  \psi -\psi (\nabla^{\mu}+ i q A^{\mu}) \psi^{*}\Bigg].
\end{eqnarray}
In the limiting case where  $b\rightarrow 0$, the above field
equations reduce to the equations of holographic superconductor in
Maxwell theory \cite{Har2}. We take the following metric ansatz
for a planar black hole with backreaction \cite{Har2}
\begin{eqnarray}\label{metric}
ds^{2}&=- f(r) e^{-\chi(r)} dt^{2}+\frac{dr^2}{f(r)}+r^{2}(dx^2 +
dy^2),
\end{eqnarray}
where $f(r)$ and $\chi(r)$ are functions of $r$ only. The
electromagnetic field and the scalar field can be chosen as
\cite{Har}
\begin{eqnarray}\label{phipsi}
A_{\mu}=\left(\phi(r),0,0,0\right)   \  \   \   \psi=\psi(r).
\end{eqnarray}
It is worth noting that because of gauge freedom, we can choose
$\psi$ to be a real scalar field. The Hawking temperature of the
black hole, which will be interpreted as the temperature of the
CFT, is given by
\begin{eqnarray}\label{T}
T=\frac{f'(r_{+}) e^{-\chi(r_{+})/2}}{4\pi},
\end{eqnarray}
where the prime denotes the derivative with respect to $r$, and
$r_{+}$ is the black hole horizon defined by $f(r_{+})=0$.
Inserting the metric (\ref{metric}) as well as (\ref{phipsi}) in
the field equations (\ref{Eq1})-(\ref{Eq3}), the components of the
Einstein equations lead to
\begin{eqnarray}
\chi'&+2 r \kappa^{2} \left(\psi'^{2}+\frac{q^{2}e^{\chi} \phi^{2}
\psi^2}{f^2}\right)=0, \label{Eqchi}
\end{eqnarray}
\begin{eqnarray}
f^{\prime}-\left(\frac{3 r}{L^{2}}-\frac{f}{r}\right)+ r\kappa^{2}
\left[m^{2} \psi^{2}-\frac{1}{b} +\frac{1}{b \sqrt{1-b e^{\chi}
\phi'^{2}}} +f\left(\psi '^{2}+\frac{q^{2} e^{\chi}
\phi^{2}\psi^{2}}{f^{2}}\right)\right]=0,\label{Eqf}
\end{eqnarray}
while the scalar and gauge field equations become
\begin{eqnarray}
\phi^{\prime\prime} +\frac{2}{r} \phi^{\prime} \left(1-b \phi'^{2}
e^{\chi}\right)+\frac{\phi' \chi'}{2} -\frac{2 q^{2} \psi^{2} }{f}
\phi\left(1-b {\phi ^{\prime}}^{2} e^{\chi} \right)^{{3}/{2}}
=0,\label{Eqphi}
\end{eqnarray}
\begin{eqnarray}
\psi^{\prime\prime} +\left(\frac{2}{r}-\frac{\chi
^{\prime}}{2}+\frac{f'}{f}\right) \psi' -\frac{m^{2}}{f} \psi+
\frac{q^{2}\phi^{2} e^{\chi}}{f^{2}} \psi=0.\label{Eqpsi}
\end{eqnarray}
One may note that for $b \rightarrow 0$, the field equations
(\ref{Eqchi}) -(\ref{Eqpsi}) restore the field equations of
holographic superconductor with backreaction in Maxwell theory
\cite{PJWC2012}, as expected. In this paper, we would like to
consider the backreaction of the bulk fields on the background
metric that describes a charged Born-Infeld black hole in the AdS
bulk. We re-scale the bulk fields $\phi$, $\psi$ and the
Born-Infeld coupling parameter $b$ as $\phi\rightarrow\phi/q$,
$\psi \rightarrow\psi/q$ and $b\rightarrow q^2 b$. Under these
transformations, the form of the gauge and the scalar field
equations do not change, but the gravitational coupling in the
Einstein equation changes $\kappa^2\rightarrow \kappa^2/q^2$. In
general the probe limit is defined as $\kappa^2/q^2\rightarrow0$.
There are two methods to include the backreaction of matter fields
on the metric. The first method is to consider $\kappa^2=1$ and
choose a finite value of $q^2$ as described in \cite{Har2}. In
this approach, the probe limit is equivalent to letting
$q\rightarrow \infty$ \cite{PJWC2012}.  In the second method, one
can fix $q^2=1$ \cite{ssym} and consider finite values of the
parameter $\kappa^2$. In this case the probe limit corresponds to
letting $\kappa^2\rightarrow 0$. In this paper, we adopt the
second approach to fix the backreaction parameter to be
$\kappa^2$. Therefore, in the limiting case where
$\kappa^2\rightarrow 0$, the field equations restore those of the
holographic superconductors in Born-Infeld electrodynamics in the
probe limit \cite{Gan}.

For the normal phase where $\psi(r)=0$, from  Eq. (\ref{Eqchi}) we
find that $\chi$ is a constant and the metric becomes the
Reissner-Nordstr\"{o}m AdS black hole as the Born-Infeld nonlinear
parameter $b$ approaches to zero. Thus, we have
\begin{eqnarray} \label{frphr}
f(r)=\frac{r^2}{L^2}-\frac{1}{r}
\left(\frac{r_{+}^3}{L^2}+\frac{\kappa^2 \rho^2}{2
r_{+}}\right)+\frac{\kappa^2 \rho^2}{2 r^2},  \   \   \
\phi\approx\mu-\frac{\rho}{r},
\end{eqnarray}
where $\mu$ and $\rho$ are interpreted as the chemical potential
and charge density in the holographic superconductor \cite{Har}.
Note that when the Born-Infeld factor is not equal to zero, the
solution is the Born-Infeld-AdS black hole.

Since we are interested in getting solution for superconducting
phase where $\psi\neq0$, we must impose the appropriate boundary
conditions. At the black hole horizon, $r_{+}$, we have
$f(r_{+})=0$ and the regularity conditions $\phi(r_{+})=0$ for the
gauge field \cite{reg}, imply the boundary conditions
\begin{eqnarray}
\psi(r_{+})=\frac{f'(r_{+})\psi'(r_{+})}{m^2},
\end{eqnarray}
and the coefficients in the metric ansatz satisfy
\begin{eqnarray}
\chi'(r_{+})&=&-2 \kappa^2 r_{+}\left(\psi'^2+
\frac{e^{\chi(r_{+})}\phi'(r_{+})^2
\psi(r_{+})^2}{f'(r_{+})^2}\right),\\\label{f+&chi+}
f'(r_{+})&=&\frac{3r_{+}}{L^2}- \kappa^2 r_{+} \left
[m^2\psi(r_{+})^2-\frac{1}{b}+\frac{1}{b\sqrt{1-b\phi'^{2}(r_{+})
e^{\chi(r_{+})}}} \right].
\end{eqnarray}
Far from the horizon boundary, at the spatial infinity where
$r\rightarrow\infty$, the asymptotic performance of the solutions
are
\begin{eqnarray}\label{B.C}
\chi\rightarrow0,
 \  \   f\approx\frac{r^2}{L^2},  \  \  \  \phi\approx\mu-\frac{\rho}{r},  \
 \  \
 \psi\approx\frac{\psi_{-}}{r^{\Delta_{-}}}+\frac{\psi_{+}}{r^{\Delta_{+}}},
\end{eqnarray}
where
\begin{eqnarray}\label{Delta}
\Delta_{\pm}=\frac{3\pm\sqrt{9+4 m^2}}{2}.
\end{eqnarray}
According to the gauge/gravity duality, $\psi$ can be regarded as
the source of the dual operator $\mathcal{O}$,
$\psi_{-}=<\mathcal{O_{-}}>$  and  $\psi_{+}=<\mathcal{O_{+}}>$,
respectively. Setting $m^2=-2$ in (\ref{Delta}), we have
$\Delta_{-}=1$ and $\Delta_{+}=2$. Following \cite{Har,Har2}, we
can impose the boundary condition in which either $\psi_{+}$ or
$\psi_{-}$ vanishes, so that the theory is stable in the
asymptotic AdS region. In the following calculation, we will focus
on the condition $\psi_{+}=0$. Moreover, we will consider the
values of $m^2$ which must satisfy the Breitenlohner-Freedman (BF)
bound $m^2\geq-9/4$ \cite{BF} for the $4$-dimensional spacetime.
In the remaining part of this paper we will set $L=1$.
\section{Analytical Investigation of the Holographic Superconductor\label{Field}}
In this section, we would like to study the $(2+1)$-holographic
superconductor phase transition in the presence of Born-Infeld
nonlinear electrodynamics by taking into account the backreaction
of the scalar and gauge field on the metric background. We employ
the Sturm-Liouville variational method and investigate the
relation between the critical temperature of condensation and the
charge density near the phase transition point. In particular, we
shall examine the effects of the backreaction as well as the
Born-Infeld parameters on the critical temperature. In order to
solve Eqs. (\ref{Eqchi})-(\ref{Eqpsi}), we rewrite them in terms
of a new dimensionless coordinate, $z={r_{+}}/{r}$. The result is
\begin{eqnarray}\label{chiz}
\chi^{\prime}-2 \kappa^{2}\left(z
{\psi^{\prime}}^{2}+\frac{r_{+}^{2}}{z^{3} f^{2}} e^{\chi}
\phi^{2} \psi^{2}\right)=0,
\end{eqnarray}
\begin{eqnarray}\label{fz}
f^{\prime}-\frac{f}{z} +\frac{3 r_{+}^{2}}{L^{2} z^{3}} -
\frac{\kappa^{2} r_{+}^{2}}{z^{3}}\left[m^{2} \psi^{2}
-\frac{1}{b}+\frac{1}{b}\frac{1}{\sqrt{1-b e^{\chi}
\frac{z^{4}}{r_{+}^{2}}{\phi^{\prime}}^{2}}}+ f
\left(\frac{z^{4}}{r_{+}^{2}} {\psi^{\prime}}^{2}+\frac{1}{f^{2}}
e^{\chi} \phi^{2} \psi^{2}\right)\right]=0,
\end{eqnarray}
\begin{eqnarray}\label{phiz}
\phi^{\prime\prime}+\frac{\phi^{\prime} \chi^{\prime}}{2} +
\frac{2b e^{\chi} z^{3} }{r_{+} ^{2}} {\phi^{\prime}}^{3} -\frac{2
r_{+}^{2} \psi^{2} }{z^{4} f} \phi \left(1-\frac{b e^{\chi} z^{4}
}{r_{+} ^{2}} {\phi^{\prime}}^{2} \right)^{3/2}=0
\end{eqnarray}
\begin{eqnarray}\label{psiz}
\psi^{\prime\prime}-\left(\frac{\chi^{\prime}}{2}
-\frac{f^{\prime}}{f}\right) \psi^{\prime} - \frac{r_{+}^{2}}{z^4}
\left(\frac{m^2}{f}-\frac{e^\chi \phi^2}{f^2}\right) \psi=0
\end{eqnarray}
where the prime now indicates the derivative with respect to $z$.
In the absence of the backreaction, the solution of Eq. (\ref{fz})
is
\begin{eqnarray}
f(z)=r_{+}^2 \left(\frac{1}{z^2}-z\right),
\end{eqnarray}
and Eqs. (\ref{phiz}) and (\ref{psiz}) reduce to their
corresponding equations in Ref. \cite{Gan}. In the vicinity of the
critical temperature, $T_{c}$, which the stability is confirmed
\cite{instability}, we can select the order parameter as an
expansion parameter because it has a small value \cite{PJWC2012}
\begin{eqnarray}
\epsilon\equiv <\mathcal{O}_{i}>,
\end{eqnarray}
with $i=+$ or $i=-$. Since we are interested in solutions where
$\psi(r)$ is small, therefore from Eqs. (\ref{phiz}) and
(\ref{psiz}) we can expand the scalar field $\psi$ and the gauge
field $\phi$ as \cite{SK}
\begin{eqnarray}
\psi=\epsilon \psi_{1}+\epsilon^3 \psi_{3}+\epsilon^5\psi_{5}+...,
\end{eqnarray}
\begin{eqnarray}
\phi=\phi_{0}+\epsilon^2\phi_{2}+\epsilon^4\phi_{4}+...,
\end{eqnarray}
where $\epsilon\ll1$. The metric functions $f(z)$ and $\chi(z)$
can also be expanded around the Reissner-Nordstr\"{o}m AdS
spacetime
\begin{eqnarray}
f=f_{0}+\epsilon^2 f_{2}+\epsilon^4f_{4}+...,
\end{eqnarray}
\begin{eqnarray}
\chi=\epsilon^2\chi_{2}+\epsilon^4\chi_{4}+....
\end{eqnarray}
For the chemical potential $\mu$, we allow it to be expanded as the following series form
\begin{eqnarray}
\mu=\mu_{0}+\epsilon^2\delta \mu_{2}+...,
\end{eqnarray}
where $\delta \mu_{2}>0$. Thus, near the phase transition, the
order parameter  as a function of the chemical potential can be
obtained as
\begin{eqnarray}
\epsilon\thickapprox\Bigg(\frac{\mu-\mu_{0}}{\delta
\mu_{2}}\Bigg)^{1/2}.
\end{eqnarray}
It is clear that when $\mu$ approaches $\mu_{0}$, the order
parameter $\epsilon$ approaches zero. The phase transition occurs
at the critical value $\mu_{c}=\mu_{0}$. Note that the critical
exponent $\beta=1/2$ is the universal result from the
Ginzburg-Landau mean field theory. At the zeroth order, the
equation of motion for $\phi$ reduces to
\begin{eqnarray}
\phi^{\prime\prime}(z) +
\frac{2bz^3}{r_{+c}^2}{\phi^{\prime}}^{3}(z)=0.
\end{eqnarray}
If we set $\phi^{\prime}(z)=\xi (z)$, we have
\begin{eqnarray}\label{xi0}
\xi^{\prime}(z) + \frac{2bz^3}{r_{+c}^2}\xi^3 (z)=0.
\end{eqnarray}
Integrating the above equation in the interval $[0,1]$, yields
\begin{eqnarray}\label{xip}
\frac{1}{\xi^2(1)}-\frac{1}{\xi^2(0)}=\frac{b}{r^2_{+c}},
\end{eqnarray}
where $\xi=\xi(0)$ at $z=0$ and $\xi=\xi(1)$ at $z=1$, and from
Eq. (\ref{frphr}) we have
\begin{eqnarray}\label{phip}
\phi^{\prime}(0)=\xi(0)\approx
-\frac{\rho}{r_{+}}=-\frac{\rho}{r_{+c}},
\end{eqnarray}
at $T=T_c$. Also, from Eqs.  (\ref{xip}) and (\ref{phip}), we
obtain
\begin{eqnarray}\label{xip1}
\frac{1}{\xi^2(1)}=\frac{b}{r^2_{+c}}+\left(\frac{r_{+c}}{\rho}\right)^2.
\end{eqnarray}
Integrating Eq. (\ref{xi0}) in the interval $[1,z]$, after using
Eq. (\ref{xip1}) we arrive at
\begin{eqnarray}\label{xi1}
\xi(z)=\phi^{\prime}(z)=-\frac{\lambda r_{+c}}{\sqrt{1+b\lambda^2
z^4}},
\end{eqnarray}
where we have taken the negative sign in the expression for
$\phi^{\prime}(z)$ since $\phi^{\prime}(0)$ is negative at $z = 0$
and
\begin{eqnarray}\label{lambda}
\lambda=\frac{\rho}{r^2_{+c}}.
\end{eqnarray}
Integrating Eq. (\ref{xi1}) from $z^{\prime}=1$ to $z^{\prime}=z$,
we get
\begin{eqnarray}
\phi(z)=-\int_{1}^{z} \frac{\lambda r_{+c}}{\sqrt{1+b \lambda^2
{z^{\prime}}^{4}}} dz^{\prime},
\end{eqnarray}
where we have used the fact that $\phi(z=1)=0$. Since the above
integral cannot be solved exactly, we shall expand the integrand
binomially up to $\mathcal{O} (b)$. We find
\begin{eqnarray}\label{phi0}
\phi_{0}(z)= \lambda r_{+c} (1-z)\Bigg( {1- \frac{b \lambda^2}{10}
(1+z+z^2+z^3+z^4)}\Bigg), \  \   \  b\lambda^2< 1.
\end{eqnarray}
At the zeroth order, the equation for $f$ has the following
solution
\begin{eqnarray}
f_{0}(z) = r_{+} ^2 g(z)= r_{+} ^2 \left[ \frac{1}{z^2} -z -
\frac{\kappa ^2 \lambda^2}{2} z(1-z) + \frac{b}{40} \kappa^2
\lambda^4 z(1-z^5)\right],
\end{eqnarray}
where we introduce the new function $g(z)$ for simplicity in the
following calculations. It is worth noting that we shall assume
the deviation from the linear Maxwell field is small. This allows
us to keep only the nonlinear parameter $b$ up to the first order.
Now, in the first order approximation, the asymptotic AdS boundary
conditions $(z\rightarrow 0)$ for $\psi$ can be expressed as
\begin{eqnarray}
\psi_{1} \approx \frac{\psi_{-}}{r_{+}^{\Delta_{-}}}
z^{\Delta_{-}}+ \frac{\psi_{+}}{r_{+}^{\Delta_{+}}}z^{\Delta_{+}}.
\end{eqnarray}
In order to match the behavior at the boundary, we can define
\begin{eqnarray}\label{psi1F}
\psi_{1}(z)= \frac{<\mathcal{O}_i>}{ \sqrt{2
}r_{+}^{\triangle_{i}}} z^ {\triangle_{i}}F(z),
\end{eqnarray}
where $F(z)$ is a trial function near the boundary $z=0$  which
satisfies the boundary conditions $F(0)=1$ and $F'(0)=0$
\cite{Sio}. Inserting Eq. (\ref{psi1F}), we can write Eq.
(\ref{psiz}) as
\begin{eqnarray}
&&F^{\prime\prime}(z)+\Bigg[\frac{2\Delta}{z}+\frac{g'}{g}\Bigg]
F'(z)+\Bigg[\frac{\Delta}{z}\Bigg(\frac{\Delta-1}{z}+
\frac{g'}{g}\Bigg)-\frac{m^2}{z^4g}\Bigg]
F(z)\nonumber\\&&+\frac{\lambda^2 (1-z)^2}{z^4 g^2}\Bigg[1-\frac{b
\lambda^2}{5} \Bigg(1+z+z^2+z^3+z^4\Bigg)\Bigg]F(z)=0.
\end{eqnarray}
This equation can be rewritten as a Sturm-Liouville eigenvalue equation
\begin{eqnarray}
T F^{\prime\prime}+T' F' + P F +\lambda^2 Q F=0,
\end{eqnarray}
where $T$, $P$, and $Q$ read
\begin{eqnarray}
T(z)&=&z^{2 \Delta_{i}+1} \Bigg[2(z^{-3} -1)-\kappa^2
\lambda^2(1-z)+\frac{b}{20}\kappa^2 \lambda^4 (1-z^5)\Bigg],\\
P(z)&=& \frac{\Delta_{i}}{z}\Bigg(\frac{\Delta_{i}-1}{z}+ \frac{g'}{g}\Bigg)-\frac{m^2}{z^4g},\\
Q(z)&=& \frac{ (1-z)^2}{z^4 g^2}\Bigg[1- \frac{b \lambda^2}{5}
\Bigg(1+z+z^2+z^3+z^4\Bigg)\Bigg].
\end{eqnarray}
According to the boundary conditions for $F(z)$, we can take the
trial function as
\begin{eqnarray}
F(z)=1-\alpha z^2.
\end{eqnarray}
The minimal eigenvalue $\lambda^2$ is obtained by minimizing the
following expression with respect to the coefficient $\alpha$
\begin{eqnarray}
\lambda^2=\frac{\int_{0}^{1} T\left(F'^2-P
F^2\right)dz}{\int_{0}^{1} TQF^2  dz}.
\end{eqnarray}
In order to simplify the following calculation, we will express
the backreacting parameter as \cite{LPJW2015}
\begin{eqnarray}
\kappa_{n}= n \Delta\kappa,   \    \   \    n=0,1,2,...
\end{eqnarray}
where $\Delta\kappa=\kappa_{n+1} - \kappa_{n}$ is the step size of
our iterative procedure. We are interested in finding the effects
of the nonlinear corrections on the backreaction term, i.e. we
want to obtain the $\lambda^2$ up to the order $\kappa^2$,
\begin{eqnarray}
\kappa^2 \lambda^2= \kappa_{n} ^2 \lambda^2=\kappa_{n} ^2
(\lambda^2|_{\kappa_{n-1}})
+\mathcal{O}\left[(\Delta\kappa)^4\right].
\end{eqnarray}
Here we have set $\kappa_{-1}=0$ and $\lambda^2|_{\kappa_{-1}}=0$.
We shall perform a perturbative expansion $b \lambda^2$ and retain
only the terms that are linear in $b$ such that
\begin{eqnarray}
b \lambda^2=b \left(\lambda^2|_{b=0}\right) +\mathcal{O} (b^2),
\end{eqnarray}
where $\lambda^2|_{b=0}$ is the value of $\lambda^2$ for $b=0$. In
fact, we have only retain  the terms that are linear in
Born-Infeld parameter $b$.

According to the definition of $T$, the critical temperature
$T_{c}$ is given by
\begin{eqnarray}
T_{c}=\frac{f^{\prime} (r_{+c})}{4 \pi}.
\end{eqnarray}
Using Eq. (\ref{f+&chi+}), we have
\begin{eqnarray}
f^{\prime}(r_{+c})=3 r_{+c}-\kappa^2 r_{+c}
\left[-\frac{1}{b}+\frac{1}{b
\sqrt{1-b{{\phi_{0}}^{\prime}}^{2}(r_{+c})}} \right].
\end{eqnarray}
Thus, the critical temperature $T_{c}$ can be expressed as
\begin{eqnarray}\label{TC}
T_{c}=\frac{1}{4\pi} \sqrt{\frac{\rho}{\lambda}}\Bigg[ 3-
\frac{\kappa_{n} ^2 (\lambda^2|_{\kappa_{n-1}})}{2}+ \frac{1}{8} b
\kappa_{n}^2(\lambda^4|_{\kappa_{n-1},b=0}) \Bigg],
\end{eqnarray}
where we have used Eq. (\ref{phi0}) as well as relation
\begin{eqnarray}
b\kappa^2 \lambda^4= b \kappa_{n}^2(\lambda^4|_{\kappa_{n-1},b=0})
+\mathcal{O}(b^2)+\mathcal{O}[(\Delta\kappa)^4].
\end{eqnarray}
In this way we present a complete picture of the critical
temperature $T_{c}$ for the $(2+1)$-dimensional holographic
superconductors in Born-Infeld nonlinear electrodynamics with
backreactions. It is important to note that we shall obtain the
analytical results by taking the values of $m^2=-2$,
$\Delta_{i}=\Delta_{-}=1$ and setting $\Delta\kappa=0.05$. Also,
the nonlinear parameter is taken as $b=0,0.1, 0.2, 0.3$.
Obviously, the critical temperature $T_{c}$ depends on the
parameters $\kappa$ and $b$. As an example, we bring the details
of our calculation for $n=2$ with different values of $b$. For
$b=0$, we find
\begin{eqnarray}
\lambda^2=\frac{1.663870667 \alpha^2- 0.9960856000 \alpha +
0.9978253333}{0.7153474994-0.1752999949 \alpha +0.03033207562
\alpha^2},
\end{eqnarray}
whose minimum is $1.2660$ at $\alpha=0.23813$. According to Eq.
(\ref{TC}), we get the critical temperature $T_{c}=0.2246
\sqrt{\rho}$, which is in good agreement with the result of
\cite{PJWC2012}. For $b=0.1$, we have
\begin{eqnarray}
\lambda^2=\frac{0.997816-0.996076\alpha+1.66389\alpha^2}{0.688014-0.1650222\alpha+0.0283792\alpha^2},
\end{eqnarray}
which attains its minimum $1.314677$ at $\alpha=0.239498$ and the
critical temperature reads $0.2225 \sqrt{\rho}$. For $b=0.2$, we
arrive at
\begin{eqnarray}
\lambda^2=\frac{0.997763-0.995993\alpha+1.66384\alpha^2}{0.660686-0.154746\alpha+0.0261691\alpha^2},
\end{eqnarray}
whose minimum is $1.36718$ at $\alpha=.24091$ and the critical
temperature becomes $0.2203 \sqrt{\rho}$. For $b=0.3$, we find
\begin{eqnarray}
\lambda^2=\frac{0.997455-0.995439\alpha+1.66349\alpha^2}{0.633356-0.144469\alpha+0.239587\alpha^2},
\end{eqnarray}
which has a minimum value $1.42378$ at $\alpha=0.242346$, and we
can easily get the critical temperature $0.2180\sqrt{\rho}$. Let
us summarize our results in table $1$.  From this table, we see
that, for a fixed value of the nonlinear parameter $b$, the value
of the critical temperature decreases with increasing the
backreaction parameter $\kappa$. Similar behavior between the
Born-Infeld parameter $b$ with the critical temperature has also
been observed. Namely, for fixed value of the backreaction
parameter, the critical temperature decreases with increasing the
nonlinear parameter $b$. Note that we have taken $\kappa_{n}=n
\Delta \kappa$ where $\Delta \kappa=0.05$. In general, the
presence of both Born-Infeld and backreaction decrease the
critical temperature and make the condensation harder. The
critical temperature $T_{c}$ obtained here for $\kappa=b=0$,
agrees with the analytical result of Ref. \cite{P.ZGJZ} and
numerical result in Ref. \cite{Har}. Also, considering the effect
of $b$, without the backreaction parameter $\kappa$ i.e., the
probe limit, our results are consistent with those obtained in
Ref. \cite{Gan}. On the other hand in the absence of nonlinear
electrodynamics ($b=0$), our analytical results show a good
agrement with the analytical and numerical results obtained in
Ref. \cite{PJWC2012} for the holographic superconductor with
backreaction.
\begin{center}
\begin{tabular}{|c|c|c|c|c|}
\hline
$\kappa_{n}$ \quad &   $b=0$\quad &   $b=0.1$\quad  &   $b=0.2$\quad  &   $b=0.3$\quad \\
\hline
$0$ \quad &   0.2250 \quad &   0.2228\quad  &   0.2206\quad  &   0.2184\quad \\
\hline
$0.05$ \quad&   0.2249\quad &   0.2227\quad &   0.2204\quad &    0.2181\quad \\
\hline
$0.10$ \quad &   0.2246\quad &   0.2225\quad  &   0.2203\quad  &   0.2180\quad \\
\hline
$0.15$ \quad &   0.2241\quad &   0.2220\quad  &   0.2199\quad  &   0.2176\quad \\
\hline
$0.20$ \quad&   0.2235\quad &   0.2214\quad &   0.2192\quad &   0.2170\quad \\
\hline
$0.25$ \quad &   0.2226\quad &   0.2208\quad  &   0.2184\quad  &   0.2162\quad \\
\hline
$0.30$ \quad &   0.2216\quad &   0.2196\quad  &   0.2174\quad  &   0.2152\quad \\
\hline
\end{tabular}
\\[0pt]
Table $1$: The critical temperature $T_{c}/\sqrt{\rho}$ for
different values of both parameters $b$ and $\kappa_{n}$.
\label{tab1}
\end{center}
\section{Critical Exponent and Condensation Values}\label{Cri}
In this section, we would like to compute the condensation values
of the condensation operator $\langle \mathcal{O} \rangle$ in the
boundary field theory and near the critical temperature. First of
all, we write the field equation (\ref{phiz}) by using Eq.
(\ref{psi1F}) in the form
\begin{eqnarray}\label{phiCT}
\phi^{\prime\prime} + \frac{2b z^{3} }{r_{+} ^{2}}
{\phi^{\prime}}^{3}= \frac{\langle \mathcal{O} \rangle ^2}{r_{+}
^2}  \mathcal{B} (z) \phi (z),
\end{eqnarray}
\begin{eqnarray}
\mathcal{B} (z)= \frac{F^2 (z)}{1-z^3}
\left(1-\frac{3bz^4}{2r_{+}^2} \phi'^2 (z) \right) \left[
1+\frac{\kappa^2 z^3}{1+z+z^2}
\left(\frac{\lambda^2}{2}-\frac{b\lambda^4}{40}
\xi(z)\right)\right],
\end{eqnarray}
where $\xi(z)=1+z+z^2+z^3+z^4$. Without the backreaction, this
equation reduces to the Eq. (42) of Ref.\cite{Gan}. We shall
assume the parameter ${\langle \mathcal{O} \rangle ^2}/{r_{+} ^2}$
is small. Expanding $\phi (z)$ for the small parameter ${\langle
\mathcal{O} \rangle ^2}/{r_{+} ^2}$, we get
\begin{eqnarray}\label{phibast}
\frac{\phi(z)}{r_{+}}=\lambda (1-z) \Bigg[
1-\frac{b\lambda^2}{10}\xi(z) \Bigg]+\frac{\langle \mathcal{O}
\rangle ^2}{r_{+} ^2} \chi (z).
\end{eqnarray}
With the help of Eq. (\ref{phibast}), Eq.( \ref{phiCT}) becomes
\begin{eqnarray}
\chi''(z)+6b \lambda^2 z^3 \chi'(z)=\frac{\lambda F^2}{1+z+z^2}
\Bigg[1-\frac{b\lambda^2}{10}(\xi(z)+15 z^4) \Bigg] \nonumber \\
 +\frac{\lambda F^2 z^3}{(1+z+z^2)^2} \left[ \frac{\kappa^2 \lambda^2}{2}-\frac{b\kappa^2 \lambda^4}{40} ( 3\xi(z)+30
 z^4)\right],
\end{eqnarray}
with $\chi(1)=0=\chi'(1)$. Multiplying this equation by factor
$\exp\left(\frac{3b\lambda^2 z^4}{2}\right)$, we arrive at
\begin{eqnarray}
\frac{d}{dz} \left(e^{\frac{3b\lambda^2 z^4}{2}} \chi'(z) \right)
&=&\lambda e^{\frac{3b\lambda^2 z^4}{2}} \frac{F^2}{1+z+z^2}
\left[1-\frac{b\lambda^2}{10} \left(\xi(z)+15 z^4\right) \right.
\nonumber
\\
&& \left.+\frac{z^3}{1+z+z^2} \left(\frac{\kappa^2 \lambda^2}{2} -
\frac{b \kappa^2 \lambda^4}{40} (3\xi(z)+30z^4) \right) \right].\\
\nonumber
\end{eqnarray}
Integrating both sides of the above equation between $z=0$ to
$z=1$, yields
\begin{eqnarray}
\chi'(0)=-\lambda \int_{0}^{1} dz e^{\frac{3b\lambda^2 z^4}{2}}
\frac{F^2}{1+z+z^2} \Bigg\{ 1-\frac{b\lambda^2}{10}(\xi(z)+15
z^4)+\frac{z^3}{1+z+z^2} \left[ \frac{\kappa^2 \lambda^2}{2} -
\frac{b \kappa^2 \lambda^4}{40} (3\xi(z)+30z^4) \right] \Bigg\}.
\end{eqnarray}
Combining Eq. (\ref{B.C}) with Eq. (\ref{phibast}), we have
\begin{eqnarray} \label{murho}
\frac{\mu}{r_{+}}-\frac{\rho}{r_{+} ^2} z &=& \lambda (1-z)
\Bigg\{ 1-\frac{b\lambda^2}{10} \xi(z) \Bigg\}+\frac{\langle
\mathcal{O} \rangle ^2}{r_{+} ^2} \chi (z)\nonumber\\&=& \lambda
(1-z) \Bigg\{ 1-\frac{b\lambda^2}{10} \xi(z) \Bigg\}+\frac{\langle
\mathcal{O} \rangle ^2}{r_{+} ^2} \left(\chi(0)+z
\chi'(0)+...\right).
\end{eqnarray}
Comparing the coefficient of $z$ on both sides of the Eq.
(\ref{murho}), we get
\begin{eqnarray}
\frac{\rho}{r_{+} ^2}=\lambda-\frac{\langle \mathcal{O} \rangle
^2}{r_{+} ^2}\chi'(0).
\end{eqnarray}
Substituting $\chi'(0)$ in the above equation, we reach
\begin{eqnarray}
\frac{\rho}{r_{+} ^2}=\lambda \Bigg\{ 1+\frac{\langle \mathcal{O}
\rangle ^2}{r_{+} ^2} \mathcal{A}\Bigg\},
\end{eqnarray}
where
\begin{eqnarray}
\mathcal{A}=\int_{0}^{1} dz e^{\frac{3b\lambda^2 z^4}{2}}
\frac{F^2}{1+z+z^2} \Bigg\{ 1-\frac{b\lambda^2}{10}(\xi(z)+15
z^4)+\frac{z^3}{1+z+z^2} \left[\frac{\kappa^2 \lambda^2}{2} -
\frac{b \kappa^2 \lambda^4}{40} (3\xi(z)+30z^4) \right] \Bigg\}.
\end{eqnarray}
Next, we should compute an expression for $r_{+}$. Considering the
fact that $T$ is very close to $T_{c}$ and using Eqs. (\ref{T}),
(\ref{f+&chi+}) and (\ref{phi0}), we have
\begin{eqnarray}\label{r+}
r_{+}=\frac{4\pi T}{\left[3-\frac{\kappa^2
\lambda^2}{2}+\frac{b}{8} \kappa^2 \lambda^4\right]}.
\end{eqnarray}
Finally, with the help of Eqs. (\ref{lambda}), (\ref{TC}), and
(\ref{r+}), we get the following expression,
\begin{eqnarray}
T_{c}^{2}-T^2= \langle \mathcal{O} \rangle ^2
\frac{\mathcal{A}}{(4\pi)^2} \left[3-\frac{\kappa^2
\lambda^2}{2}+\frac{b}{8} \kappa^2 \lambda^4 \right]^2.
\end{eqnarray}
Therefore, we find
\begin{eqnarray}
\langle \mathcal{O} \rangle=\gamma T_{c} \sqrt{1-\frac{T}{T_{c}}},
\end{eqnarray}
where
\begin{eqnarray}
\gamma=\frac{4\pi\sqrt{2}}{\sqrt{\mathcal{A}}}\left[3-\frac{\kappa^2
\lambda^2}{2}+\frac{b}{8} \kappa^2 \lambda^4 \right]^{-1}.
\end{eqnarray}
Thus, near the critical point, the condensation operator $\langle
\mathcal{O} \rangle$ will satisfy
\begin{eqnarray}
\langle \mathcal{O} \rangle \sim \sqrt{1-\frac{T}{T_{c}}},
\end{eqnarray}
which holds for various values of both Born-Infeld and
backreaction parameters. Also, the critical exponent is identical
to the mean field value $1/2$, which implies that the existence of
the mentioned parameters do not have any consequence on the
second-order phase transition. Considering only the Born-Infeld
parameter, the values of $\mathcal{A}$ is in good agreement with
the one in Ref. \cite{Gan}. We summarize the our results in table
$2$. We see that the condensation value increases as the
Born-Infeld parameter $b$ increases for the fixed parameter
$\kappa$. On the other hand, for fixed value of $b$  the
condensation parameter increases with increasing the backreaction
parameter. Thus, in both cases the condensation value increases,
which shows that the higher Born-Infeld electrodynamics and
gravitational backreaction corrections make the condensation to be
harder.
\begin{center}
\begin{tabular}{|c|c|c|c|c|}
\hline
$\kappa_n$ \quad &   $b=0$\quad &   $b=0.1$\quad  &   $b=0.2$\quad  &   $b=0.3$\quad \\
\hline
$0.0$ \quad &   8.07 \quad &   8.18\quad  &   8.31\quad  &   8.47\quad \\
\hline
$0.05$ \quad&   8.09\quad &   8.187\quad &   8.321\quad &   8.489\quad \\
\hline
$0.10$ \quad &   8.11\quad &   8.19\quad  &   8.324\quad  &   8.4893\quad \\
\hline
$0.15$ \quad &   8.115\quad &   8.21\quad  &   8.34\quad  &   8.50\quad \\
\hline
$0.20$ \quad&   8.13\quad &   8.24\quad &   8.37\quad &   8.53\quad \\
\hline
$0.25$ \quad &   8.16\quad &   8.29\quad  &   8.39\quad  &   8.56\quad \\
\hline
$0.30$ \quad &   8.20\quad &   8.31\quad  &   8.44\quad  &   8.60\quad \\
\hline
\end{tabular}
\\[0pt]
Table 2: The values of the condensation parameter $\gamma$ for
different values of $b$ and $\kappa_n$. \label{tab1}
\end{center}
\section{Conclusions\label{Con}}
Employing the Sturm-Liouville eigenvalue problem, we have
analytically investigated the effects of the Born-Infeld nonlinear
gauge field on the properties of $(2+1)$-dimensional holographic
supercondoctors in the background of AdS black holes. We performed
our analysis away from the probe limit, where the scalar and gauge
fields back react on the background metric. First, we presented a
detailed analysis of solving the coupled equations of motion  for
the scalar and gauge fields. We obtained the relationship between
the critical temperature and the charge density. We performed our
calculations up to the first order in the Born-Infeld coupling
parameter and up to order $\kappa^2$ in the backreaction
parameter. We observed that both backreaction and Born-Infeld
parameters make the critical temperature of the holographic
superconductor smaller. This implies that the condensation
formation is affected by both the backreaction and the Born-Infeld
coupling parameters. That is to say, the condensation becomes
harder in the presence of the backreaction and Born-Infeld
parameters. We also found that the critical exponent of the
condensation is $1/2$ which is the universal value in the mean
field theory. The results obtained in this paper consist with the
previous numerical and analytical results in the limiting cases
where either the backreaction or the nonlinear parameters are
turned off. We expect our analytical results to be confirmed
numerically in the near future investigations. It is also
interesting to extend this investigation to other type of
nonlinear electrodynamics such as exponential, logarithmic and
power-Maxwell Lagrangian. These issues are now under
investigations and the results will be appeared soon.
\acknowledgments{We thank Shiraz University Research Council. This
work has been supported financially by Research Institute for
Astronomy and Astrophysics of Maragha (RIAAM), Iran.}

\end{document}